\documentclass[conference,a4paper]{IEEEtran}

\usepackage[totalwidth=480pt, totalheight=714pt]{geometry}

\ifCLASSINFOpdf
  \usepackage[pdftex]{graphicx}
\else
  \usepackage[dvips]{graphicx}
\fi

\ifCLASSOPTIONcompsoc
 \usepackage[caption=false,font=normalsize,labelfont=sf,textfont=sf]{subfig}
\else
 \usepackage[caption=false,font=footnotesize]{subfig}
\fi

\IEEEoverridecommandlockouts
\usepackage[T1]{fontenc}
\usepackage{blkarray, bigstrut}
\usepackage{algorithm,algorithmic}
\usepackage{colortbl}
\usepackage{pdfpages}

\usepackage{cite}
\usepackage{amsmath,amssymb,amsfonts,mathtools}
\usepackage{algorithmic}
\usepackage{graphicx}
\usepackage{textcomp}
\usepackage{xcolor}
\usepackage{hyperref}
\usepackage{siunitx}
\usepackage{tikz,tikz-3dplot}
\usepackage{pgfplots}
\usepackage{balance}
\pgfplotsset{compat=1.17}
\usepackage{pgfkeys}
\usetikzlibrary{3d}
\usetikzlibrary{patterns}
\usetikzlibrary{patterns.meta}
\usetikzlibrary{decorations.shapes}
\usetikzlibrary{fit}
\usetikzlibrary{decorations.pathreplacing}
\usetikzlibrary{decorations.markings}
\usetikzlibrary{intersections}

\colorlet{grey}{black!60}

\definecolor{green}{RGB}{17, 184, 36}

\tikzset{point/.style={draw,circle,thick,inner sep=2pt}}
\tikzstyle{ground}=[fill,draw=none,minimum width=0.3,minimum height=0.6]

\tikzstyle{image-line}=[->,densely dashdotted,grey, very thick]
\tikzstyle{intersection-line}=[dotted,thick]
\tikzstyle{intersection-point}=[point,blue]
\tikzstyle{valid-line}=[->,green,very thick]
\tikzstyle{block}=[draw=black,thick, align=center, minimum height=1.2cm,fill=red!80!black,text=white]

\tdplotsetmaincoords{80}{45}
\tdplotsetrotatedcoords{-90}{180}{-90}

\tikzset{surface/.style={draw=black!70, fill=black!40!white, fill opacity=.6}}
\tikzstyle{wall-surface}=[blue,fill opacity=.5,draw=black,thick,dashed]

\newcommand{\coneback}[4][]{
    \draw[canvas is xy plane at z=#2, #1] (-135-#4:#3) arc (-135-#4:-245+#4:#3) -- (O) --cycle;
}
\newcommand{\conefront}[4][]{
    \draw[canvas is xy plane at z=#2, #1] (45-#4:#3) arc (45-#4:-135+#4:#3) -- (O) --cycle;
}

\newcommand{\bs}{\boldsymbol}
\newcommand{\bb}{\mathbb}
\newcommand{\cl}{\mathcal}

\newcommand{\iid}{%
  \ifmmode
  \mathrm{i.i.d.}%
  \else%
  i.i.d.\@\xspace%
  \fi%
}
\newcommand{\scp}[3][]{#1\langle #2, #3 #1\rangle}
\newcommand{\ie}{\emph{i.e.}, }
\newcommand{\eg}{\emph{e.g.}, }

\usepackage{soul} 

\def\BibTeX{{\rm B\kern-.05em{\sc i\kern-.025em b}\kern-.08em
T\kern-.1667em\lower.7ex\hbox{E}\kern-.125emX}}

\makeatletter
\newcommand\fs@spaceruled{\def\@fs@cfont{\bfseries}\let\@fs@capt\floatc@ruled
  \def\@fs@pre{\vspace{.5\baselineskip}\hrule height.8pt depth0pt \kern2pt}%
  \def\@fs@post{\kern2pt\hrule\relax\vspace{-.5\baselineskip}}%
  \def\@fs@mid{\kern2pt\hrule\kern2pt}%
  \let\@fs@iftopcapt\iftrue}
\makeatother

\usepackage{fancyhdr}
\begin{document}


\def\method{MPT}

\title{Min-Path-Tracing: A Diffraction Aware Alternative to Image Method in Ray Tracing}

\author{\IEEEauthorblockN{
Jérome Eertmans\IEEEauthorrefmark{1},   
Claude Oestges\IEEEauthorrefmark{1},   
Laurent Jacques\IEEEauthorrefmark{1},    
}                                     
\IEEEauthorblockA{\IEEEauthorrefmark{1}
(UCLouvain): Electronical Engineering, ICTEAM, Louvain-la-Neuve, Belgium,
    \href{mailto:jerome.eertmans@uclouvain.be}{firstname.lastname@uclouvain.be}}
}

\maketitle
\thispagestyle{fancy}

\begin{abstract}
    For more than twenty years, Ray Tracing methods have continued to improve on both accuracy and computational time aspects. However, most state-of-the-art image-based ray tracers still rely on a description of the environment that only contains planar surfaces. They are also limited by the number of diffractions they can simulate. We present Min-Path-Tracing (\method{}), an alternative to the image method that can handle diffractions seamlessly, while also leveraging the possibility to use different geometries for surfaces or edges, such as parabolic mirrors. \method{} uses implicit representations of objects to write the path finding challenge as a minimization problem. We further show that multiple diffractions can be important in some situations, which \method{} is capable to simulate without increasing neither the computational nor the implementation complexity. 
\end{abstract}


\section{Introduction}\label{sec:intro}
Over the past decades, Ray Tracing (RT) has gained increased interest in computer graphics~\cite{marrsRayTracingGems2021} and telecommunication fields~\cite{yunRayTracingRadio2015,espostiRayTracingTechniques2021}. Generally speaking RT's goal is to compute every possible path between two nodes, and later apply appropriate physical wave propagation rules to determine a channel model for communication between those nodes, \eg between base station (BS) and user equipment (UE), and derive some important metrics, such as the path loss or interference level. A variety of RT implementations can be found, either with deterministic outcomes (\eg Image RT) or stochastic (\eg Ray Launching). However, regarding image based RT, modern ray-tracers often suffer from limitations on both the number of diffractions and the type of geometries they can handle, \ie mostly polygons \cite{heDesignApplicationsHighPerformance2019}.

In this paper, we describe Min Path Tracing (\method{}), an alternative to the image method (IM) that allows us to generalize the \emph{path finding} process, \ie the computation of all possible paths between two nodes, regardless of the geometries of 3-D scene or the number of diffractions encountered along the path. Our technique leverages, if available, the implicit equations of surfaces and edges in the scene to construct a minimization problem. Then, the paths coordinates are obtained as solutions of this problem. The structure of this work is organized as follows. First, we define necessary mathematical tools and notations: Sec.~\ref{sec:problem} establishes the problem we are solving, and Sec.~\ref{sec:path_candidates} describes how to generate the set of all possible lists of interactions. Next, we detail how IM (Sec.~\ref{sec:image}) and our alternative method (Sec.~\ref{sec:alt}) work in practice. In Sec.~\ref{sec:algo}, we summarize the main steps of the computation of paths between the BS and UE nodes in a single algorithm. Then, in Sec.~\ref{sec:application}, we compute electric field contributions from different paths, in a simple urban scenario, to highlight the importance of intermediate diffraction in radiocommunications. Finally, we conclude our work in Sec.~\ref{sec:ccl} by comparing both methods and discussing the future applications of \method{}.

\section{Problem definition}\label{sec:problem}

A key part of any RT technique is the path finding step. This step aims to determine one or more paths between two nodes, \eg BS and UE, that undergo multiple interactions with the environment. In the frame of this paper, we restrict our analysis to reflections and diffractions, and we assume that we know, for each facet, an implicit equation $f(x,y,z)=0$ whose $(x,y,z)$-solutions include the facet's coordinates. In other words, we first suppose to have infinite surfaces, and we will later consider their actual frontier. Additionally, we assume that we also have an implicit equation for each edge, as well as their direction vector at each point. The uniqueness, or existence, of a reflected or diffracted path depends on the shape of the objects the path interacts with. For simple diffraction or reflection on infinite planar surfaces and straight edges, this path is unique. However, specular reflection on concave paraboloids introduces symmetry and multiple possible solutions. Once a path is found, it must be validated. Indeed, as we first assume that objects are possibly infinitely long, we can find a path with an interaction point that does not fall inside the actual object, as expected. This separation between path finding and path validation helps us to develop methods that are agnostic of the object's size. In the context of this paper, the path validation step is performed a posteriori.

Let $n_t$ be the number of interactions with the environment, and $\cl L := \{L_1, \ldots, L_{n_t}\}$ the list of $n_t$ surfaces or edges. The number of reflections and diffractions are noted, respectively, $n_r$ and $n_d$, such that $n_t = n_r + n_d$. The list order matters, as the $k$-th interaction will be on the $k$-th element $L_k$ in $\cl L$. In a 3-D space, the problem of finding such path reduces to determining the location of the $n_t$ points or $3 n_t$ unknowns, one for each object in $\cl L$.

    \subsection{Specular Reflection}
    
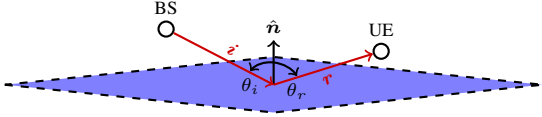
\begin{figure}[t]
        \scriptsize
        \centering
        \begin{tikzpicture}[tdplot_main_coords,yscale=.6]
            \coordinate (O) at (0,0,0);

            \coordinate (BS) at (-2,0,1);
            \coordinate (UE) at (2,0,1);
            \pgfmathsetmacro{\length}{5}
            \coordinate (L) at (-{\length/2},-{\length/2},0);

            \node[point,label=above:{BS}] (nBS) at (BS) {};
            \node[point,label=above:{UE}] (nUE) at (UE) {};

            \fill[wall-surface] (L) -- ([shift={(\length,0,0)}]L) -- ([shift={(\length,\length,0)}]L) -- ([shift={(0,\length,0)}]L) -- cycle;

            \draw[->,thick] (O) -- ++(0,0,1) node[above] {$\hat{\bs n}$};
            \draw[->,thick,red!80!black] (nBS) -- (O) node[midway,above,sloped] {$\bs i$};
            \draw[->,thick,red!80!black] (O) -- (nUE) node[midway,below,sloped] {$\bs r$};

            \draw[->,thick,canvas is xz plane at y=0] (0,.5) arc (90:150:.5) node[below] {$\theta_i$};
            \draw[->,thick,canvas is xz plane at y=0] (0,.5) arc (90:30:.5) node[below] {$\theta_r$};

        \end{tikzpicture}
        \caption{Illustration of the specular reflection, for which an incident vector reflects on a surface such that both the incident and reflected vectors make the same angle, i.e., $\theta_i=\theta_r$, with the surface normal $\hat{\bs n}$ defined in \eqref{eq:normal_vector}.}
        \label{fig:reflection}
    \end{figure}

    Specular reflection is the regular, mirror-like reflection observed when an incident wave reflects into a ray that makes the same angle with the normal vector to the surface, but from the opposite side (Fig.~\ref{fig:reflection}). Therefore, denoting vectors in bold symbols, the incident vector $\bs i$ and reflected vector $\bs r$ are related by
    \begin{equation}\label{eq:reflection}
        \hat{\bs r} = \hat{\bs \imath} - 2 \scp{\hat{\bs \imath}}{\hat{\bs n}}\hat{\bs n},
    \end{equation}
    where the vector normalization allows for arbitrary sized $\bs r$ vectors. Above, the operators $\scp{\cdot}{\cdot}$ and $\hat{\cdot}$ refer to the dot product and the normalized vector, respectively.

    Note that the surface does not have to be planar; we only need to know its local normal vector at every location in the 3-D scene. Moreover, if we possess an implicit equation of our surface, $f(x,y,z)=0$, then the normal vector can be easily derived with
    \begin{equation}\label{eq:normal_vector}
        \hat{\bs{n}} = \frac{\bs \nabla f}{\|\bs \nabla f\|},
    \end{equation}
    where $\bs \nabla$ is the gradient operator.

    \subsection{Edge Diffraction}
    
\begin{figure}[t]
        \scriptsize
        \centering
        \begin{tikzpicture}[tdplot_main_coords,yscale=.6]
            \coordinate (O) at (0,0,0);
            \pgfmathsetmacro{\zCone}{3}
            \pgfmathsetmacro{\rCone}{2}

            \pgfmathsetmacro{\angleBS}{-135}
            \pgfmathsetmacro{\zBS}{-2}
            \pgfmathsetmacro{\rBS}{abs(\zBS * \rCone / \zCone)}
            \pgfmathsetmacro{\xBS}{\rBS * cos(\angleBS)}
            \pgfmathsetmacro{\yBS}{\rBS * sin(\angleBS)}

            \pgfmathsetmacro{\angleUE}{160}
            \pgfmathsetmacro{\zUE}{3}
            \pgfmathsetmacro{\rUE}{abs(\zUE * \rCone / \zCone)}
            \pgfmathsetmacro{\xUE}{\rUE * cos(\angleUE)}
            \pgfmathsetmacro{\yUE}{\rUE * sin(\angleUE)}

            \coordinate (BS) at (\xBS,\yBS,\zBS);
            \coordinate (UE) at (\xUE,\yUE,\zUE);
            \pgfmathsetmacro{\length}{3}
            \pgfmathsetmacro{\lengthback}{1.2*\length}
            \pgfmathsetmacro{\lengthx}{0.5*\lengthback}
            \coordinate (L) at (0,0,-{\length/2});

            \node[point,label=below:{BS}] (nBS) at (BS) {};
            \node[point,label=above:{UE}] (nUE) at (UE) {};

            \coneback[surface]{3}{2}{0}
            \fill[wall-surface] (L) -- ([shift={(-\lengthx,\lengthback,0)}]L) -- ([shift={(-\lengthx,\lengthback,\length)}]L) -- ([shift={(0,0,\length)}]L) -- cycle;

            \draw[->,thick,red!80!black] (nBS) -- (O) node[midway,above,sloped] {$\bs i$};
            \draw[->,thick,red!80!black] (O) -- (nUE) node[midway,above,sloped,rotate=180] {$\bs d$};

            \fill[wall-surface] (L) -- ([shift={(\length,0,0)}]L) -- ([shift={(\length,0,\length)}]L) -- ([shift={(0,0,\length)}]L) -- cycle;

            \draw[->,thick] (0,0,-2.5) -- (0,0,4) node[above] {$\hat{\bs e}$};
            \conefront[surface]{\zCone}{\rCone}{0}

            \draw[thick,dashed] (nBS) -- (0,0,\zBS) node[midway,above,sloped] {$h_i$};
            \draw[thick,dashed] (nUE) -- (0,0,\zUE) node[midway,above,sloped,shift={(0,0,.2)}] {$h_d$};

            \pgfmathsetmacro{\dx}{0.2}
            \draw[thick] (0,0,\zBS) -- ++({\dx * \xBS},{\dx * \yBS},0) -- ++({-\dx * \xBS},{-\dx * \yBS},{\dx*\rBS}) -- cycle;
            \draw[thick] (0,0,\zUE) -- ++({\dx * \xUE},{\dx * \yUE},0) -- ++({-\dx * \xUE},{-\dx * \yUE},{\dx*\rUE}) -- cycle;

        \end{tikzpicture}
        \caption{Illustration of the Keller diffraction cone, which gives rise to a multitude of diffracted rays, such that $h_i/\|\bs i \| = h_d /\|\bs d\|$, equivalent to \eqref{eq:diffraction}. One of those rays creates a path from BS to UE.}
        \label{fig:diffraction}
    \end{figure}
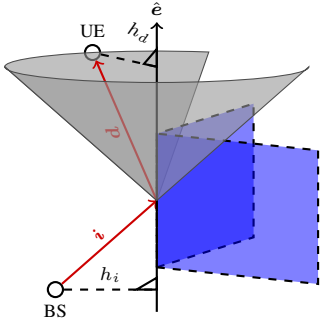

    For electromagnetic (EM) waves with sufficiently high frequency, i.e., when the size of the scatterer is large when compared to the wavelength, we can approximate the diffraction phenomena using Keller's diffraction theory \cite{kellerGeometricalTheoryDiffraction1962}. The latter states that an incident vector $\bs i$ diffracts on an edge with local direction vector $\hat{\bs e}$ into a multitude of rays $\bs d$ that build up a cone such that $\bs i$ and $\bs d$ make the same angle with $\hat{\bs e}$ (Fig.~\ref{fig:diffraction}). Incident and diffracted vectors respect therefore this relation:
    \begin{equation}\label{eq:diffraction}
        \frac{\scp{\bs i}{\hat{\bs e}}}{\| \bs i \|} =  \frac{\scp{\bs d}{\hat{\bs e}}}{\|\bs d\|}.
    \end{equation}
    
    If one knows a parametric equation of the edge $\bs r(s)$ for some parametrization $s\in \bb R$, then the direction vector is simply equal to its derivative:
    \begin{equation}
        \hat{\bs{e}} = \frac{\bs{r}'(s)}{\|\bs{r}'(s)\|},
    \end{equation}
    with $\bs{r}'(s) = \mathrm{d}\bs{r}/\mathrm{d}s$.

\section{Finding Path Candidates}\label{sec:path_candidates}
    
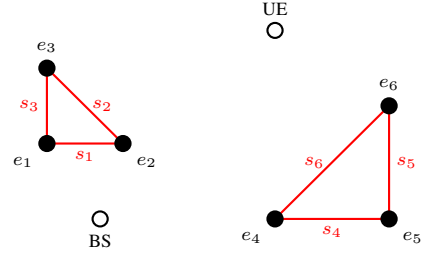
\begin{figure}[t]
        \scriptsize
        \centering
        \begin{tikzpicture}
            \node[point,label=below:{BS}] (BS) at (-.3,-.5) {};
            \node[point,label=above:{UE}] (UE) at (2,2) {};

            \draw[red,thick] (-1,.5) node[point,black,fill=black,label=south west:{\color{black}$e_1$}] {}
            -- ++(1,0) node[midway,below] {$s_1$} node[point,black,fill=black,label=south east:{\color{black}$e_2$}] {}
            -- ++(-1,1) node[midway,right] {$s_2$} node[point,black,fill=black,label=above:{\color{black}$e_3$}] {}
            -- cycle node[midway,left] {$s_3$};

            \draw[red,thick] (2,-.5) node[point,black,fill=black,label=south west:{\color{black}$e_4$}] {}
            -- ++(1.5,0) node[midway,below] {$s_4$} node[point,black,fill=black,label=south east:{\color{black}$e_5$}] {}
            -- ++(0,1.5) node[midway,right] {$s_5$} node[point,black,fill=black,label=above:{\color{black}$e_6$}] {}
            -- cycle node[midway,left] {$s_6$};
        \end{tikzpicture}
        \caption{2-D scenario with triangular-shaped objects on which reflection or diffraction can occur. Surfaces are colored in \textcolor{red}{red} and edges in black.}
        \label{fig:visibility}
    \end{figure}

    Both IM and \method{} require a list of interactions $\cl L$ to estimate the ray path. In our approach, we consider a graph-based construction to deduce the interaction list of each possible path. From the visibility matrix of the scene (\eg Fig. \ref{fig:visibility}), an adjacency matrix is built (see Fig. \ref{eq:adjacency}) so that it represents a directed graph that encodes the list of all possibles trajectories going from BS to UE.

    We now describe IM, before developing our alternative approach, the \method{} method.

    \begin{figure}[htbp]
    \scriptsize
    \begin{equation*}
        \makeatletter\setlength\BA@colsep{4pt}\makeatother
        \begin{blockarray}{ccccccccccccccc}
            & \text{BS} & s_1 & s_2 & s_3 & s_4 & s_5 & s_6 & e_1 & e_2 & e_3 & e_4 & e_5 & e_6 & \text{UE}\\
            \begin{block}{c(>{\medspace}c|cccccc|cccccc|c<{\medspace})}
                \text{BS} & {} & 1 & {} & {} & {} & {} & 1 & 1 & 1 & {} & 1 & {} & 1 & 1\\
                \BAhhline{~~------------~}
                s_1 & {} & {} & {} & {} & {} & {} & \bs{1} & {} & {} & {} & \bs{1} & {} & & {} & {}\\ 
                s_2 & {} & {} & {} & {} & {} & {} & \bs{1} & {} & {} & {} & \bs{1} & {} & \bs{1} & 1\\ 
                s_3 & {} & {} & {} & {} & {} & {} & {} & {} & {} & {} & {} & {} & {} & {}\\ 
                s_4 & {} & {} & {} & {} & {} & {} & {} & {} & {} & {} & {} & {} & {} & {}\\ 
                s_5 & {} & {} & {} & {} & {} & {} & {} & {} & {} & {} & {} & {} & {} & {}\\ 
                s_6 & {} & \bs{1} & \bs{1} & {} & {} & {} & {} & \bs{1} & \bs{1} & \bs{1} & {} & {} & {} & 1\\ 
                e_1 & {} & {} & {} & {} & {} & {} & \bs{\color{red}1} & {} & {} & {} & \bs{\color{red}1} & {} & {} & {}\\ 
                e_2 & {} & {} & {} & {} & {} & {} & \bs{\color{red}1} & {} & {} & {} & \bs{\color{red}1} & {} & \bs{\color{red}1} & 1\\ 
                e_3 & {} & {} & {} & {} & {} & {} & \bs{\color{red}1} & {} & {} & {} & \bs{\color{red}1} & {} & \bs{\color{red}1} & 1\\ 
                e_4 & {} & \bs{\color{red}1} & \bs{\color{red}1} & {} & {} & {} & {} & \bs{\color{red}1} & \bs{\color{red}1} & \bs{\color{red}1} & {} & {} & {} & 1 \\ 
                e_5 & {} & {} & {} & {} & {} & {} & {} & {} & {} & {} & {} & {} & {} & {}\\ 
                e_6 & {} & {} & \bs{\color{red}1} & {} & {} & {} & {} & {} & \bs{\color{red}1} & \bs{\color{red}1} & {} & {} & {} & 1\\
                \BAhhline{~~------------~}
                \text{UE} & {} & {} & {} & {} & {} & {} & {} & {} & {} & {} & {} & {} & {} & {}\\
            \end{block}
        \end{blockarray}
    \end{equation*}
    \caption{Adjacency matrix, $\mathcal{G}$, generated from scenario illustrated on Fig. \ref{fig:visibility}. Each row of this $14\times14$ matrix refers to the visible objects as seen from the corresponding object. For readability purposes, zeros are discarded. Inside $\mathcal{G}$, one can find the visibility matrix, $\mathcal{V}$, whose coefficients are highlighted in {\bf bold}. In the case of IM, only part of this matrix is used. If one uses a similar method to \cite{quatresoozTrackingInteractionPoints2021} that allows for diffraction at last interaction, the coefficients in \textcolor{red}{red} would be discarded from $\mathcal{G}$, which dramatically reduces the number of path candidates from BS to UE.}
    \label{eq:adjacency}
    \end{figure}

\subsection{The Image Method}\label{sec:image}

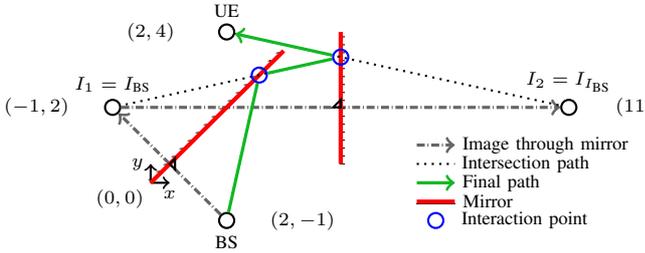
\begin{figure}[htbp]
        \scriptsize
        \centering
        \begin{tikzpicture}[scale=.5]

            \draw[->,black,thick] (0,0) -- ++(.5,0) node[below] {$x$};
            \draw[->,black,thick] (0,0) -- ++(0,.5) node[left] {$y$};
            \node[anchor=north east] at (0,0) {$(0,0)$};

            \node[point,label=below:{BS},name=BS] (BS) at (2,-1) {};
            \node[point,label=above:{UE},name=UE] (UE) at (2,4) {};

            \node[right of=BS] {$(2, -1)$};
            \node[left of=UE] {$(2, 4)$};

            \node[point,label=above:{$I_1 = I_\text{BS}$},name=I1] (I1) at (-1,2) {};
            \draw[image-line] (BS) -- (I1);
            \node[point,label=above:{$I_2 = I_{I_\text{BS}}$},name=I2] (I2) at (11,2) {};
            \draw[image-line] (I1) -- (I2);

            \node[right of=I2] {$(11,2)$};
            \node[left of=I1] {$(-1,2)$};

            \draw[ground,pattern={Lines[angle=0]},rotate around={45:(0,0)}] (0,0) rectangle ({3.5/cos(45)}, .1);
            \draw[name path=W1,red,ultra thick] (0,0) -- (3.5,3.5);
            \draw[ground,pattern={Lines[angle=45]}] (5,.5) rectangle (5.1, 4);
            \draw[name path=W2,red,ultra thick] (5,.5) -- (5,4);

            \draw[intersection-line,name path=L1] (UE) -- (I2);
            \path[name intersections={of=W2 and L1,by=PR1}];
            \node[intersection-point] (R1) at (PR1) {};
            \draw[intersection-line,name path=L2] (R1) -- (I1);
            \path[name intersections={of=W1 and L2,by=PR2}];
            \node[intersection-point] (R2) at (PR2) {};
            \draw[intersection-line] (R2) -- (BS);
            \draw[valid-line] (BS) -- (R2) -- (R1) -- (UE);

            \pgfmathsetmacro{\shift}{0.0}
            \pgfmathsetmacro{\dx}{0.2}
            \pgfmathsetmacro{\stwo}{sqrt(2)*0.5}
            \path[name path=L1] (BS) -- (I1);
            \path[name intersections={of=W1 and L1,by=IM1}];
            \draw[thick] ([shift={(\shift,0)}]IM1) -- ++ ({\dx*\stwo},-{\dx*\stwo}) -- ++(0,{2*\dx*\stwo}) -- cycle;
            \path[name path=L2] (I1) -- (I2);
            \path[name intersections={of=W2 and L2,by=IM2}];
            \draw[thick] ([shift={(-{\stwo*\shift},{\stwo*\shift})}]IM2) -- ++ (-\dx,0) -- ++(\dx,\dx) -- cycle;

            \begin{scope}[shift={(7,1,0)}]
                \draw[image-line] (0,0) -- ++(1,0) node[right,black] {Image through mirror};
                \draw[intersection-line] (0,-.5) -- ++(1,0) node[right,black] {Intersection path};
                \draw[valid-line] (0,-1) -- ++(1,0) node[right,black] {Final path};
                \draw[ground,pattern={Lines[angle=45]}] (0, -1.5) rectangle ++(1,-.1);
                \draw[red,ultra thick] (0,-1.5) -- ++(1,0) node[right,black] {Mirror};
                \path (0,-2) -- ++(1,0) node[right] {Interaction point} node[midway,intersection-point] {};
            \end{scope}
        \end{tikzpicture}
        \caption{Example application of IM in RT. The method determines the only valid path that can be taken to join BS and UE with, in between, reflection with two mirrors (the interaction order is important). First, the consecutive images of the BS are determined through each mirror, using line symmetry. Second, intersections with mirrors are computed backward, \ie from last mirror to first, by joining the UE, then the intersections points, with the images of the BS. Finally, the valid path can be obtained by joining BS, the intermediary intersection points, and the UE.}
        \label{fig:image_method}
    \end{figure}

    IM determines the exact paths between BS and UE, with a certain number of specular reflections, by computing the successive images of the BS by orthogonal symmetries on surfaces. As illustrated in Fig.~\ref{fig:image_method}, all images are first computed successively through each surface: the BS image through the first surface is computed, then the image of this image using the second surface, and so on until the last surface is reached. This forward pass is summarized in the following equation:
    \begin{equation}
        \bs{I}_k = \bs{I}_{k-1} - 2 \scp[\big]{\bs I_{k-1} - \bs P_k}{\hat{\bs n}_k}\hat{\bs n}_k ,\label{eq:reflection_forward}
    \end{equation}
    with $\bs I_k$ and $\bs P_k$, respectively, the $k$-th image and any point on the $k$-th surface, and $\bs I_0 = \text{BS}$.
    
    Next, the interaction points are computed, from last to first, by determining the intersection of each surface and the path joining the previous point, or the UE, and the corresponding image:
    \begin{equation}
        \bs{X}_k = \bs{X}_{k+1} + \frac{\scp[\big]{\bs P_k - \bs X_{k+1}}{\hat{\bs n}_k}}{\scp[\big]{\bs{X}_{k+1} - \bs{I}_{k} }{\hat{\bs n}_k}} \Big(\bs{X}_{k+1} - \bs{I}_{k} \Big),\label{eq:reflection_backward}
    \end{equation}
    with $\bs X_k$ the interaction point on the $k$-th surface, $\bs X_0 = \text{BS}$, and $\bs X_{n_t+1} = \text{UE}$.
    
    This, however, is only valid for reflections on planar surfaces. To account for diffraction, different approaches exist, such as using an analytical solution and only allowing one diffraction to occur at the last interaction \cite{quatresoozTrackingInteractionPoints2021}. Handling diffraction with IM introduces non-trivial implementations and often leads to discarding most of them. As explained hereafter, our method aims at developing a low-complexity implementation of diffraction while also allowing for non-planar geometries.

\subsection{Min-Path-Tracing method}\label{sec:alt}

     In the Min-Path-Tracing (MPT) method, we express the path finding problem as a minimization program enforcing the estimated path to satisfy both \eqref{eq:reflection} and \eqref{eq:diffraction}. For each reflection (resp. diffraction), we assume to know the normal vector to the surface (resp. the direction vector to the edge), at all points in the space.
     
     As \eqref{eq:diffraction} requires normalized vectors, we rewrite the reflection equation \eqref{eq:reflection}, as
    \begin{equation}\label{eq:reflection_modified}
        \gamma \cdot \bs r = \bs i - 2 \scp{\bs i}{\hat{\bs n}}\hat{\bs n},
    \end{equation}
    with $\gamma = \| \bs i \|/\| \bs r \|$.

    The path finding problem has to determine $3n_t = 3n_r + 3n_d$ unknowns. Defining the points $\bs X_0$ and $\bs X_{n_t+1}$ as the BS and UE locations, respectively, each of the $n_t$ interactions depends on $9$ unknowns, three 3-D points, namely the point of departure $\bs X_{k-1}$ from the previous interaction, the interaction point $\bs X_{k}$ on $L_k \in \cl L$, and the point of arrival at the next interaction, $\bs X_{k+1}$. Accordingly, \eqref{eq:reflection_modified} and \eqref{eq:diffraction} can be rewritten as 
    $$
    \cl I^{\rm r}_k(\bs X_{k-1},\bs X_{k},\bs X_{k+1}) = \bs 0, \quad \cl I^{\rm d}_k(\bs X_{k-1},\bs X_{k},\bs X_{k+1}) = 0,
    $$
    respectively, with the functions
\begin{align}
\label{eq:reflection-unknowns}
    &\begin{multlined}[t] \cl I^{\rm r}_k(\bs X_{k-1},\bs X_{k},\bs X_{k+1}) := 
    \gamma_k \cdot (\bs X_{k+1} - \bs X_k)\\
    - \big( (\bs X_{k} - \bs X_{k-1})- 2 \scp[\big]{\bs X_{k} - \bs X_{k-1} }{\hat{\bs n}_k}\,\hat{\bs n}_k \big)\end{multlined}\\
\label{eq:diffraction-unknowns}
    &\begin{multlined}[t] \cl I^{\rm d}_k(\bs X_{k-1},\bs X_{k},\bs X_{k+1}) :=\\
    \textstyle \qquad \frac{\scp[\big]{\bs X_{k} - \bs X_{k-1}}{\hat{\bs e}_k}}{\| \bs X_{k} - \bs X_{k-1} \|} - \frac{\scp[\big]{\bs X_{k+1} - \bs X_k }{\hat{\bs e}_k}}{\|\bs X_{k+1} - \bs X_k \|}.
    \end{multlined}
\end{align}
Note that in practice the functions $\cl I^{\rm r}_k$ and $\cl I^{\rm d}_k$ can be rescaled to avoid singular denominators in \eqref{eq:reflection-unknowns} and \eqref{eq:diffraction-unknowns}.

If we want to find all $n_t$ points $\bs{\cl X} := \{\bs X_k\}_{k=1}^{n_t}\in \bb R^{3n_{t}}$ that satisfy the above equations, we can equivalently find the root of the vector function $\mathcal{I}: \bb R^{n_t} \to \bb R^{3n_r + n_d}$ defined as
\begin{multline}
    \cl I(\bs{\cl X}) = \\
    \big[ \mathcal{I}_1 (\bs X_0, \bs X_1, \bs X_2)^\top,\, \ldots, \mathcal{I}_{n_t} (\bs X_{n_t-1},\bs X_{n_t},\bs X_{n_t+1})^\top\big]^\top,\label{eq:implicit_i}
\end{multline}
where the function $\mathcal{I}_k$ is either $\cl I^{\rm r}_k$ or $\cl I^{\rm d}_k$, depending on the nature of the $k$-th interaction (reflection or diffraction). 

In addition to \eqref{eq:implicit_i}, each interaction points must lie on corresponding surfaces or edges. Therefore, the implicit equations of these elements can provide the additional constraints
\begin{equation}
    f_k(\bs X_k) = 0.
\end{equation}
By introducing the function $\cl F: \bb R^{3n_t} \to \bb R^{n_t}$ with $\cl F(\bs{\cl X}) := [f_1(\bs X_1), \ldots, f_{n_t}(\bs X_{n_t})]^\top$, the equation
$$
\cl F(\bs{\cl X}) = \bs 0
$$
is satisfied if all the points lie inside their respective surfaces or edges.

Consequently, the path finding problem amounts to verifying if a minimizer $\bs{\cl X}^*$ of the optimization problem
\begin{equation}
\label{eq:minimum}
    \underset{\bs{\cl X} \in \bb R^{n_t}}{\text{minimize}}\ \cl C(\bs X) := \|\cl I(\bs X)\|^2 + \|\cl F(\bs X)\|^2,
\end{equation}
reaches a zero cost function $\cl C(\bs X)$. In this case, there exists a path corresponding to all the listed interactions. In general, the cost $\cl C$ in \eqref{eq:minimum} is not a convex, and numerous local minima can exist. However, we observed numerically that with configurations involving planar surfaces and straight edges, solving \eqref{eq:minimum} with a gradient descent converges toward the desired solutions, regardless of the initialization. For more general cases, \eg where surfaces are not necessarily planar, the minimization process should be run multiple times with different (random) initialization. Minima such that $\cl C(\bs{\cl X}) \neq 0$ can exist and should be discarded\footnote{In practice, due to numerical imprecisions, one could reach $\cl C(\bs{\cl X}^*) \neq 0$ but small anyway, while ${\cl X}^*$ is a valid solution.}.

Note that \eqref{eq:minimum} can be simplified if one knows the parametric expression of both the surfaces and edges as we can then reduce the number of unknown from $3n_t$ to $2n_r + n_d$, as surfaces and edges are fully described by two and one variables, respectively. As a result, we can use a mapping between parametric and Cartesian variables for each surface or edge, 
\begin{align}
    (s_k, t_k) &\leftrightarrow (x_k, y_k, z_k),\ \text{for surfaces},\\
    (t_k) &\leftrightarrow (x_k, y_k, z_k),\ \text{for edges},
\end{align}
so that the parametrization $\bs{\cl X}(\bs{\cl T})$, with $\bs{\cl T}$ gathering the parameters $(s_k, t_k)$ or $t_k$, directly accounts for the constraints $\|\cl F(\bs{\cl X})\|=0$. Then, \eqref{eq:minimum} boils down to solving
\begin{equation}\label{eq:minimum_param}
    \underset{\bs{\cl T} \in \bb R^{2n_r + n_d}}{\text{minimize}}\ \big\|\cl I\big(\bs{\cl  X}(\bs{\cl T})\big)\big\|^2,
\end{equation}
where the solution is now obtained in the parametric space~$\bs{\cl T}$.

This new method, minimizing \eqref{eq:minimum_param}, was validated against IM for the simplified 2-D situation depicted in Fig.~\ref{fig:image_method} and is available as a supplementary material\footnote{Access the full symbolic resolution: \url{https://tinyurl.com/symsol}.}. Indeed, MPT's path is identical to the one found previously by IM. Finally, we provide an idiomatic code\footnote{Access our method implemented: \url{https://tinyurl.com/MPTimpl}.} that implements \method{} on arbitrary geometries.

\section{Path Tracing Algorithm}\label{sec:algo}

Algorithm \ref{algo} summarizes the different steps we utilize to determine all the physically correct paths from some BS to a UE, with possibly up to $n_t$ interactions with the surrounding objects. In this algorithm, \textit{"find\_minimum\_path"} refers to the numerical solving for paths using \method{}. Lines \ref{line:vis}, \ref{line:adj}, and \ref{line:set} refer, respectively, to the construction of the visibility and adjacency matrices, and the initialization of the set that will contain all valid paths. From that, we generate the set of all path candidates, \ie the set of lists of interactions. Then, for each candidate path $p$, we run our minimizer to find the path coordinates. The \textit{"interaction\_list"} method returns the necessary information about the selected types of interactions. We repeat the minimization process $m$ times, with the value of $m$ set as a compromise between speed, robustness against local minima, and allowing for multiple solutions to \eqref{eq:minimum}.

\floatstyle{spaceruled}
\restylefloat{algorithm}
\begin{algorithm}
    \caption{Tracing paths between two nodes}
    \label{algo}
    \begin{algorithmic}[1]
    \renewcommand{\algorithmicrequire}{\textbf{Input:}}
    \renewcommand{\algorithmicensure}{\textbf{Output:}}
    \REQUIRE Maximum number of interactions $n_t$, objects database $\mathcal{D}$ , BS position and UE position
    \ENSURE List of paths from BS to UE, stored in $\mathcal{S}$
    \smallskip
    \\ \textit{Initialization}
    \STATE $\mathcal{V} \leftarrow \text{visibility\_matrix}(\mathcal{D})$\label{line:vis}
    \STATE $\mathcal{G} \leftarrow \text{adjacency\_matrix}(\text{BS},\mathcal{V},\text{UE})$\label{line:adj}
    \STATE $\mathcal{S} \leftarrow \emptyset$\label{line:set}
    \smallskip
    \\ \textit{Generate path candidates using} \href{https://networkx.org/documentation/stable/reference/algorithms/generated/networkx.algorithms.simple\_paths.all\_simple_paths.html\#networkx.algorithms.simple\_paths.all\_simple\_paths}{\texttt{NetworkX}}\textit{'s syntax}
    \STATE $\mathcal{P} \leftarrow \text{all\_simple\_paths}(G,\text{BS},\text{UE},n_t+2)$
    \smallskip
    \\ \textit{Iterate over all paths}
    \FOR {path $p$ in $\mathcal{P}$}
    \STATE $\cl L \leftarrow \text{interaction\_list}(p)$
    \STATE \textbf{repeat $m$ times}
    \begin{ALC@g}
    \STATE $\bs{\cl X}_0 \leftarrow \text{random\_guess()}$
    \STATE $\bs{\cl X}, \cl C(\bs{\cl X}) \leftarrow \text{find\_minimum\_path}(\cl L, \bs{\cl X}_0)$\label{line:min}
    \IF {($\cl C(\bs{\cl X}) <$ threshold) \textbf{and} ($\bs{\cl X}$ is valid)}\label{line:zeroeq}
    \STATE $\mathcal{S} \leftarrow \mathcal{S} \cup \{\bs{\cl X}\}$
    \ENDIF
    \end{ALC@g}
    \STATE \textbf{end repeat}
    \ENDFOR
    \end{algorithmic}
\end{algorithm}

\section{Application to an Urban Scenario}\label{sec:application}

\begin{table*}[!t]
    \centering
    \caption{Received electric field at UE, sorted by paths with similar interactions, and divided by the received field from a theoretical line of sight (LOS) path. Letter D is for diffraction and letter R is for reflection. \eg RRD stands for all the paths that encounter two reflections and one diffraction. Paths with relative contribution above \SI{-80}{\decibel} are marked in \textbf{bold}.}
    \label{tab:received_power}
    \begin{tabular}{l|r|r|r|r|r|r|r|r|r|r|r}
        Number of interactions & \multicolumn{1}{r|}{1}  & \multicolumn{3}{r|}{2} & \multicolumn{7}{r}{3} \\
        \hline
         & \multicolumn{1}{r|}{}  & \multicolumn{3}{r|}{} & \multicolumn{7}{r}{} \\
        Interaction list & D & RD & DR & DD & RRD & RDR & RDD & DRR & DRD & DDR & DDD \\
        $E/E_\text{LOS}$ (\si{\decibel}) & \textbf{-32} & -236 & -242 & \textbf{-44} & -231 & -246 & \textbf{-69} & -212 & \textbf{-72} & -81 & \textbf{-60} \\
    \end{tabular}
\end{table*}

\begin{figure}[t]
    \scriptsize
    \centering
    \begin{tikzpicture}[xscale=.8, yscale=.6]
        \draw[fill=grey!30] (0, 0) rectangle ++(1, 2);
        \draw[fill=grey!30] (3.7, 0) rectangle ++(1, 4);
        \draw[fill=grey!30] (1.8, 0) rectangle ++(1, 1);
        \draw[very thick] (-.3,0) -- (5, 0);
        \node[point,circle,label=above:{BS}] (BS) at (0.5, 2.3) {};
        \node[point,circle,label=above:{UE}] (UE) at (1.4, .4) {};
        \draw[red,dashed] (BS) -- (1,2) -- (UE);
        \draw[blue,dashed] (BS) -- (3.7,1.5) -- (1.8,1) -- (UE);
    \end{tikzpicture}
    \caption{2-D projection of the scenario of interest. In this, the LOS and most reflection paths are blocked so that diffraction becomes the principal mean of propagation for the information. Two paths (among all possible) are shown: single (in \textcolor{red}{red}) and double (in \textcolor{blue}{blue}) interaction.}
    \label{fig:geometry-basic-utd}
\end{figure}
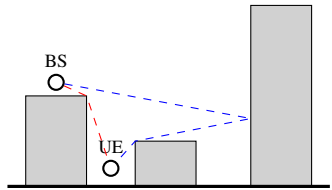

Within the frame of radiocommunications, one can combine our method with the Uniform Theory of Diffraction (UTD) to estimate, \eg the EM fields. Here, we developed a simple urban geometry with downlink communications between BS and UE where building edge diffraction plays an important role (see Fig. \ref{fig:geometry-basic-utd}). In this scenario, the BS antenna is an ideally isotropic linearly polarized antenna transmitting at \SI{1}{\giga\hertz} such that its generated electric field is
\begin{equation}
    \bs{E}(r) = \frac{E_0}{r} e^{-jkr} \hat{\bs\theta},
\end{equation}
with $r$ the distance to the observation point $Q_O$, $\hat{\bs\theta}$ the vertical polarization vector, $k$ the wavenumber, and $E_0$ the magnitude of the electric field at $r=\SI{1}{\meter}$.

For the sake of simplicity, we consider that surfaces are planar and assimilated to perfect electrical conductors. Buildings have a $\SI{15}{\meter}\times\SI{15}{\meter}$ square base and have a height of (from left to right) $y=20$, 10 and \SI{40}{\meter}. Their center is located at $x=0$, 15 and \SI{27}{\meter}. BS and UE's coordinates are, respectively, $(x,y)=(\SI{0}{\meter},\SI{22}{\meter})$ and $(\SI{8}{\meter},\SI{2}{\meter})$. After a reflection or a diffraction, the received field is, respectively,
\begin{align}
    \bs{E}^r(s) &= \bs{E}(Q_R) \cdot \overbrace{\overline{\bs{R}}\,\tfrac{r}{r+s}\, e^{-jks}}^{\overline{\bs{C}}(\text{reflection})},\\
    \bs{E}^d(s) &= \bs{E}(Q_D) \cdot \underbrace{\overline{\bs{D}}\,\sqrt{\tfrac{r}{s(r+s)}}\, e^{-jks}}_{\overline{\bs{C}}(\text{diffraction})},
\end{align}
where $Q_R$ (resp. $Q_D$) is the point of reflection (resp. diffraction), $s$ is the distance from $Q_O$ to $Q_R$ (resp. $Q_D$), $\bs{E}(Q_R)$ (resp. $\bs{E}(Q_D)$) is the received field at $Q_R$ (resp. $Q_D$), $\overline{\bs{R}}$ (resp. $\overline{\bs{D}}$) is the dyadic reflection (resp. diffraction) coefficient, and $r$ is the distance from BS to the point of interaction. More details can be found in \cite{mcnamaraIntroductionUniformGeometrical1990,paknysrobertUniformTheoryDiffraction2016}.

In general, it is well known that one cannot simply chain UTD diffraction coefficients \cite{leeGTDRayField1978} to account for multiple consecutive diffractions, but rather use specific coefficients for a given number of diffractions \cite{schneiderGeneralUniformDouble1991, carluccioUTDTripleDiffraction2012}. However, in the situation depicted in Fig. \ref{fig:geometry-basic-utd}, two consecutive diffractions are never in the transition region of each other. This is shown by the fact the transition function required to compute $\overline{\bs D}$ is always equal to one. Therefore, we are in a case where UTD reduces to the Geometrical Theory of Diffraction \cite{paknysrobertUniformTheoryDiffraction2016,kellerGeometricalTheoryDiffraction1962}, and we can apply each diffraction individually.

The total received field can be rewritten as
\begin{equation}
    \bs{E}(\text{UE}) = \sum\limits_{\bs{\cl X} \in \mathcal{S}} \bs{E}(\bs{\cl X}_1)\prod\limits_{L_k(\bs{\cl X})} \overline{\bs{C}}(L_k),
\end{equation}
where $\overline{\bs{C}}$ is the dyadic coefficient of interaction $L_k$, that accounts for reflection or diffraction depending on what applies.

Table \ref{tab:received_power} summarizes the received electrical field at UE from different levels of interaction with the environment. It shows that paths with diffraction as intermediate interaction (\eg DDD) can contribute more than paths with other types of interaction (\eg DRD). Here, single diffraction is by far the strongest path, but one could imagine scenarios where single (D) and double (DD) diffractions are blocked, meaning paths with three levels of interaction become dominant contributors to the received power.

\section{Conclusion}\label{sec:ccl}

We conclude this study by discussing the different fields of application of our method, as well as its limitations. Then, we also compare its performance with respect to IM. We finally provide a few perspectives for future developments.

\paragraph*{Fields of application}

First, for our method to work, we need a precise representation of the environment. Except for polygon-only geometries, obtaining the implicit equation for objects is not trivial, which increases the level of details required for our method to work. Besides, diffraction coefficients become increasingly hard to compute for non-planar surfaces or with lossy materials, \ie when part of the power is absorbed by the materials. In such cases, we suggest approximating the environment with a discrete representation, such as triangular meshes. This way, we could locally model surfaces as planar polygons, and extract edges from there. It is worth noting that MPT will probably not scale well with the number of objects, but so does the image method.

Second, we showed in \eqref{eq:minimum_param} that our problem greatly simplifies if we can obtain a parametric mapping for every object. In the specific case of polygon-only geometries, deriving a parametric equation from the set of points that defines a polygon or an edge can easily be done.

\paragraph*{Comparison with image method}

In terms of computational complexity, IM is linear with the number of
the number of interactions, while the complexity of our method depends on the solver that is used for the minimization, but is at least as expensive as IM.

However, the scenario presented in Fig.~\ref{fig:geometry-basic-utd} shows that double or triple diffraction can play an important role in radiocommunications, which IM cannot predict. \method{} is therefore an extension of IM that adds more possibilities on what can be simulated, in exchange to a slightly higher computational cost.

\paragraph*{Future work}

In this paper, we neglected other types of interaction than reflection or diffraction. Nevertheless, we could easily extend our method to, \eg account for refraction. Indeed, using Snell's law, we know that incident and refracted vectors are linked together. As such, adding support for refraction would amount to inserting a new equation, similar to \eqref{eq:reflection}, in our model. Diffuse scattering is also a type of interaction that can play an important role in radiocommunications, but was not studied here.

Next, we did not detail how to minimize \eqref{eq:minimum}. From \eqref{eq:reflection-unknowns} and \eqref{eq:diffraction-unknowns}, we observe that our system has a tridiagonal form, and one could use this information to accelerate the minimization process by reducing the amount of necessary computations.

\bibliographystyle{IEEEtran}
\bibliography{IEEEabrv,biblio}

\end{document}